\documentclass[prl,twocolumn,showpacs,amsmath,amssymb,superscriptaddress]{revtex4}
\usepackage{graphicx,epstopdf}
\usepackage{braket}

\newcommand{\Fkt}[1]{\,\mathsf {#1}}
\ifx\Tr\renewcommand{\Tr}{\Fkt{Tr}}
\else\newcommand{\Tr}{\Fkt{Tr}}
\fi

\begin{document}

\title{Exploiting Non-Markovianity of the Environment for Quantum Control}

\author{Daniel M. Reich}
\affiliation{Theoretische Physik, Universit\"{a}t Kassel,
  Heinrich-Plett-Str. 40, D-34132 Kassel, Germany} 
\author{Nadav Katz}
\affiliation{Racah Institute of Physics, The Hebrew University of
  Jerusalem, Jerusalem 91904, Israel}
\author{Christiane P. Koch}
\affiliation{Theoretische Physik, Universit\"{a}t Kassel,
  Heinrich-Plett-Str. 40, D-34132 Kassel, Germany} 
\email{christiane.koch@uni-kassel.de}
\date{\today}

\begin{abstract}
  When the environment of an open quantum system is non-Markovian,
  amplitude and phase
  flow not only from the system into the environment but
  also back. Here we show that this feature can be exploited to
  carry out quantum control tasks that could not be realized if the system
  was isolated. Inspired by recent experiments on superconducting
  phase circuits, we consider an anharmonic ladder with resonant amplitude
  control only. This restricts realizable operations to SO(N). The
  ladder is immersed in an environment of two-level systems. 
  Strongly coupled two-level systems lead to non-Markovian
  effects, whereas the weakly coupled ones result in
  single-exponential decay. Presence of the environment allows for
  implementing diagonal unitaries that, together with SO(N), yield the
  full group SU(N). Using optimal control theory, we obtain 
  errors that are solely $T_1$-limited. 
\end{abstract}

\maketitle

Quantum control, employing external fields to steer the outcome of a
dynamical process~\cite{RiceBook,ShapiroBook}, holds the promise of 
utilizing entanglement and matter interference as cornerstones of
future technologies. Accurate and reliable control solutions may be
identified by optimal control, provided the control target is
reachable. This question is addressed by
controllability analysis. For closed quantum systems, the answer is
determined solely by symmetries in the Hamiltonian and the available  
resources such as power or bandwidth of the controls~\cite{DAlessandroBook}.  
Controllability and control strategies for non-Markovian open quantum
systems remain largely uncharted territory. Non-Markovianity refers to
memory effects in the environment and the built-up of non-negligible
correlations between system and environment~\cite{BreuerJPB12}. It 
is generic for condensed phase settings encountered e.g. in
light harvesting or solid-state devices. Non-Markovianity can be
measured in terms of information flowing from the environment back into
the system~\cite{BreuerPRL09}, increase of correlations if the system
is bi- or multipartite~\cite{RivasPRL10}, or re-expansion of the volume of
accessible states in Liouville space~\cite{LorenzoPRA13}. Each of
these measures 
holds a promise for improved control for non-Markovian compared to
Markovian open systems: Partial recovery of coherence or growth of
correlations or a
larger accessible state space volume should all clearly facilitate
control. Indeed, correlations between system and environment 
may improve fidelities of single qubit gates~\cite{RebentrostPRL09},
cooperative effects of control and dissipation may allow for entropy
export and thus cooling~\cite{SchmidtPRL11}, and  
harnessing non-Markovianity may enhance the efficiency of quantum
information processing and communication~\cite{LaineSciRep14,BylickaSciRep14}.   

Here, we go beyond merely improving a given figure of merit 
and show that a non-Markovian environment may enable the implementation
of quantum operations that could not be realized without presence of
the environment. Our approach is based on separating the environment
into potentially beneficial and potentially detrimental parts, with the 
latter setting the timescale for (almost) unitary operations. We
employ optimal control theory (OCT) to best exploit the beneficial
non-Markovian part of the environment while beating decoherence due to
the detrimental Markovian part. For a four-level anharmonic ladder
system with resonant amplitude control only, which by itself is 
SO(4)-controllable, we demonstrate that full SU(4)-controllability
can be achieved due to the 
presence of the environment. The fidelities are only limited by the
Markovian decay. 

We investigate quantum control for a four-level system since
analytical solutions to the problem of population inversions can be
obtained by Pythagorean coupling~\cite{SuchowskiPRA11} which allow 
for realizing arbitrary operations in SO(4).
A recent experimental demonstration employed resonant amplitude control in a
flux-biased Josephson phase circuit~\cite{Svetitsky14}. 
The simplest way to construct an arbitrary element
of SU(N), provided that one is able to implement any
element of SO(N), is obtained by the Cartan decomposition.
It results in a decomposition
of all unitaries $U\in\,$SU(N) into local operations, $k_1,k_2\in\,$SO(N), 
and a diagonal, unitary matrix $A$
such that $U = k_1 A k_2$~\cite{DAlessandroBook}. 
The task to achieve full unitary controllability on the $N$-level
system therefore reduces to implementing an 
arbitrary diagonal unitary. This is the problem we address in the
following, employing OCT. 

We consider an anharmonic $N$-level system that interacts, possibly
strongly, with an environment. This interaction leads to 
(i) pure dephasing due to long-time memory,
low-frequency noise; 
(ii) energy relaxation due to weakly coupled near-resonant environmental
nodes; and (iii) visible splittings in the systems's energy levels due
to strongly coupled near-resonant 
environmental nodes. 
The strongly coupled modes are best accounted for
explicitly ("primary bath"), and we assume here that they can be
modeled by two-level systems  
(TLS)~\cite{BaerJCP97,KochJCP02,GelmanJCP04,GualdiPRA13}. Both $N$-level
system and primary bath are weakly coupled to a thermal reservoir
("secondary bath") to account for effects (i) and (ii)~\footnote{
  Strictly speaking, the low-frequency noise can also lead to
  non-Markovian effects~\cite{Galperin2006}. However, purely 
  transversal coupling to a single environmental TLS can often fully
  reproduce the experimentally observed behavior, 
  see e.g. Ref.~\cite{Lisenfeld2010}. We therefore absorb the effect
  of the low-frequency modes into the phenomenological $T_{1}$ and $T_{2}$
  times.}. 
This is modelled by a Markovian master equation ($\hbar=1$)
for the joint state of system ("Q") and primary bath ("P"), 
\begin{equation}
  \label{eq:EoM}
  \frac{d\rho_{QP}}{dt}=-i[H_{QP},\rho_{QP}] + \mathcal{L}_S(\rho_{QP})\,,
\end{equation}
with the Hamiltonian $H_{QP}$ generating the coherent evolution and
the Liouvillian $\mathcal{L}_S$  capturing the effect of the
secondary bath ("S"). The state of the system alone, $\rho_Q$, 
is obtained by integrating over the primary bath
modes~\cite{BaerJCP97,KochJCP02} that can give rise to non-Markovian
effects. For $n_P$ TLS in the primary bath, $H_{QP}$ reads 
\begin{equation}
  \label{eq:Ham}
  H_{QP} = H_Q + \sum_{i=1}^{n_P} H^{(i)}_{P} + \sum_{i=1}^{n_P}
  H_{int}^{(i)}\,,
\end{equation}
with $H_Q$ describing an 
anharmonic ladder, $E_n = n\omega_Q + \beta n(n+1)/2$, 
with base frequency $\omega_Q$ and anharmonicity $\beta$ plus 
control by an external field.
The $i$th TLS is characterized by the splitting
$\omega_i$, $H^{(i)}_P=\omega_i\sigma^{z}_{i}$, and couples
transversally to the $N$-level system,  
\begin{equation}
  \label{eq:coupling}
  H_{int}^{(i)} = \frac{S^{(i)}}{2}\left(a\sigma_{i}^+ + 
    a^\dagger\sigma_{i}^-\right)\,,
\end{equation}
with $a^+$ ($a$) the creation (annihilation) operator of the $N$-level
system, and 
the coupling constant $S^{(i)}$ corresponding to the
system's energy level splitting when on resonance with the $i$th TLS. 
The Liouvillian models decay of system and primary bath, 
\begin{equation}
  \label{eq:lindblad}
  \mathcal{L}_S (\rho) = \sum_k \left(A_k \rho A_k^{\dagger} 
    - \frac{1}{2} \left[A_k^{\dagger} A_k ,\rho \right]_+\right)\,,
\end{equation}
with $A_{n}=\sqrt{n/T_{1}}\Ket{n-1}\Bra{n}$ and 
$A_i=\sqrt{1/T_{1}^{(i)}}\sigma_{i}^-$. 
In order to limit the number of parameters, 
we restrict our model to a $T_1$-limited environment. We have verified
that it effectively captures both loss and dephasing, i.e., 
adding pure dephasing characterized by $T_2^*$  behaves, in terms of
the final fidelities, similarly to Eq.~\eqref{eq:lindblad} with
increased $T_1$. A good realization of this model is
given by superconducting circuits where the TLS
correspond to dielectric defects~\cite{MartinisPRL05} and the thermal
bath can be taken at $T=0\,$K~\footnote{
  For superconducting circuits, non-Markovian effects of purely
  dephasing $1/f$ noise are overshadowed by the strongly transversally
  coupled TLS, while the Markovian part of the $1/f$ noise leads to a
  further reduction of $T_{1}$ and $T_{2}$.}.
In particular, the TLS can be characterized experimentally in terms of
their splitting, coupling to the $N$-level system, and
$T_1$~\cite{ShaliboPRL10,ShaliboPhD}; and the upper bound of modelling 
both Markovian loss and pure dephasing by an effective $T_1$
becomes tight since 
$T_2$ is typically close to $T_2^*$~\cite{BarendsPRL13}. 

The $N$-level system is subjected to an external control $u(t)$ that
shifts its energy levels. This can be achieved, for example, by low-frequency 
steering of the bias flux in phase qudits~\cite{ShaliboPhD}. 
For low anharmonicity, the shift is harmonic, 
\begin{equation}
  \label{eq:control}
  H_{c}\left[u(t)\right] = \sum_{n=0}^{N-1} u(t)\; n \omega_Q
  \Ket{n}\Bra{n}\,. 
\end{equation}
In case of the bias flux on the phase qudit, this corresponds to
neglecting terms that oscillate strongly 
on the timescale of $\omega_Q$. 
It is those terms that, for $N=4$,
yield SO(4) operations via the Pythagorean
coupling~\cite{Svetitsky14}. 
Consequently, the two control
mechanisms, high-frequency steering on the one hand and low-frequency
steering on the other, do not interfere.
Moreover, our low-frequency control does not induce transitions to
levels with $n>4$ 
since all operators in the Hamiltonian~\eqref{eq:Ham} conserve
the occupation number of the joint state of system and primary bath. 

In the absence of the primary bath, the control
Hamiltonian~\eqref{eq:control} does not allow for realizing arbitrary
diagonal unitaries in the four-level subspace. This is best analyzed in
terms of the dynamic Lie algebra. It represents the Hilbert space
directions along which the system can evolve and is formed by nested
commutators of control and drift Hamiltonian. Since $H_c$ and $H_Q$
commute, evolution along only a single direction is possible. The
scenario changes once the strongly coupled TLS of the primary bath
come into play. In fact, a single strongly coupled TLS is sufficient
to provide the remaining $N-1$ Hilbert space directions, required for
realizing an arbitrary diagonal unitary. 
This is due to $H_{c}$ not commuting with
$H_{int}^{(i)}$. In more physical terms, $H_{int}^{(i)}$ allows for
the system wave function to be transferred to the TLS and back,  
after acquiring the desired non-local phases.

These considerations of controllability hold, however, only for
unitary evolution. The secondary bath leads to irreversible loss of
energy and phase of both system and primary bath TLS. The only control
strategy that is available for such Markovian dynamics is to beat
decoherence (unless a protected region in Hilbert space exists
in which the desired dynamics can be generated). It is thus crucial to
carry out all operations as fast as possible. Since
OCT allows for identifying controls that operate at the speed
limit~\cite{GoerzJPB11}, we use it here, employing a recent
variant for unitary gates in open quantum systems~\cite{GoerzNJP14}. 
Our optimization target is 
$U_1 = \mathrm{diag}(1,-1,1,1)$, and we quantify success in terms of
the error, $1-F_{average}$~\cite{NielsenChuang}. 
$U_1$ is a particularly difficult unitary to
implement, as exemplified by an error of over 40\%
in the absence of any strongly coupled TLS. While we discuss in the
following only $U_1$, we have verified that optimization towards
diagonal unitaries with random phases yield very similar errors~\footnote{%
  For example, optimization for 20 random diagonal unitaries, using
  the parameters of Fig.~\ref{fig:controls},  yields 
  errors between $1.390\cdot 10^{-2}$ and $1.804\cdot
  10^{-2}$, differing from that for $U_1$ by less than a
  factor of 1.2.}. This suggests full SU(4)-controllability, once
implementation of $U_1$ is successful. 

\begin{figure}[tb]
  \centering
  \includegraphics[width=0.85\linewidth]{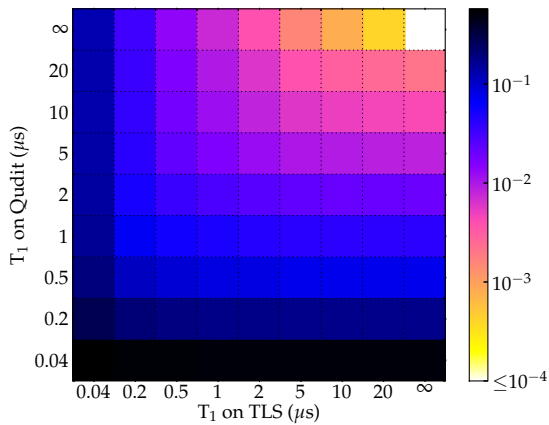}
  \caption{(color online)
    Error after optimization for $\mathrm{diag}(1,-1,1,1)$ as a function of 
    $T_1$ times of  qudit and TLS for an  optimization time of
    $T = 40\,$ns (anharmonicity $\beta = 40\,$MHz, 
    $\omega_Q - \omega^{(1)} = 550\,$MHz, $S^{(1)} = 60\,$MHz).
  }
  \label{fig:markov}
\end{figure}
Figure~\ref{fig:markov} demonstrates the interplay of Markovian and
non-Markovian effects by plotting the error for $U_1$
as a function of the $T_1$ times of qudit
and one TLS: Errors below 1\% can be reached even for $T_1$ times
of the order of a few microseconds.
Due to increasing decoherence rate with increasing 
excitation, short $T_1$ times of the qudit have a slightly more
severe effect than short $T_1$ times of the TLS. 
A multitude of controls lead to the results shown in
Fig.~\ref{fig:markov}. 
Two examples of optimized controls, obtained using different
constraints, are displayed in 
Fig.~\ref{fig:controls}(a):
\begin{figure}[tb]
  \centering
  \includegraphics[width=0.95\linewidth]{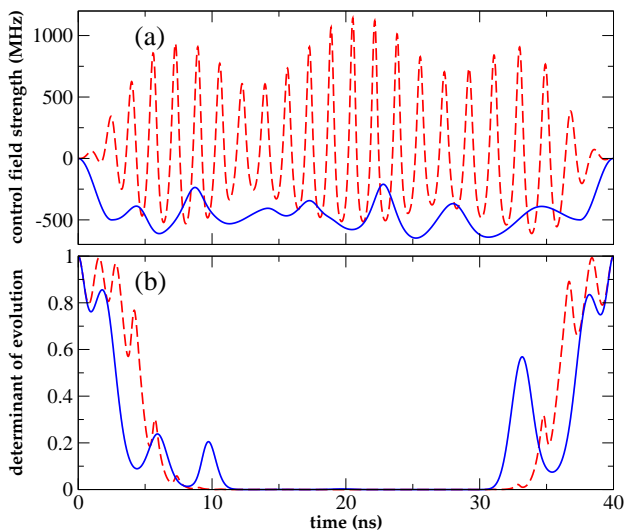}
  \caption{(color online)
    (a): Optimized amplitudes with 
    the control shown in blue following a fixed ramp of $\pm 500\,$MHz over
    $2.5\,$ns at the beginning and end and the 
    red dashed line obtained without imposing a ramp. 
    (b): Liouville space determinant of the system evolution --
    increase of the determinant indicates non-Markovianity.
  }
  \label{fig:controls}
\end{figure}
The  control can be restricted to low bandwidth by ramping it 
into and out of resonance at the beginning and end of the
optimization time interval 
(blue solid line in Fig.~\ref{fig:controls}(a)),
whereas fast oscillating controls are obtained without imposing a
ramp (red dashed line in Fig.~\ref{fig:controls}(a)). 
The different controls all share the mechanism of moving the
qudit close to resonance with the TLS, picking up a non-local phase
due to the enhanced interaction, and moving the qudit back off
resonance. This sequence is repeated several times in order to
properly align all the phases in the four-level subspace. A
visualization of the dynamics is provided as supplementary material.
While both controls lead to similar
errors, the ramped control is easier to implement experimentally
and also fulfills the low-frequency approximation
used to derive the control Hamiltonian~\eqref{eq:control}.
All further calculations therefore employ ramped controls. 
Both solutions shown in Fig.~\ref{fig:controls}(a) use 
non-Markovianity of the time evolution as core resource for control. 
This is seen in Fig.~\ref{fig:controls}(b) which plots the determinant
of the volume of reachable system states~\cite{LorenzoPRA13}, a
non-Markovianity measure that is easily evaluated numerically:
Any increase in the determinant indicates non-Markovianity. 

\begin{figure*}[tb]
  \centering
  \includegraphics[width=0.95\linewidth]{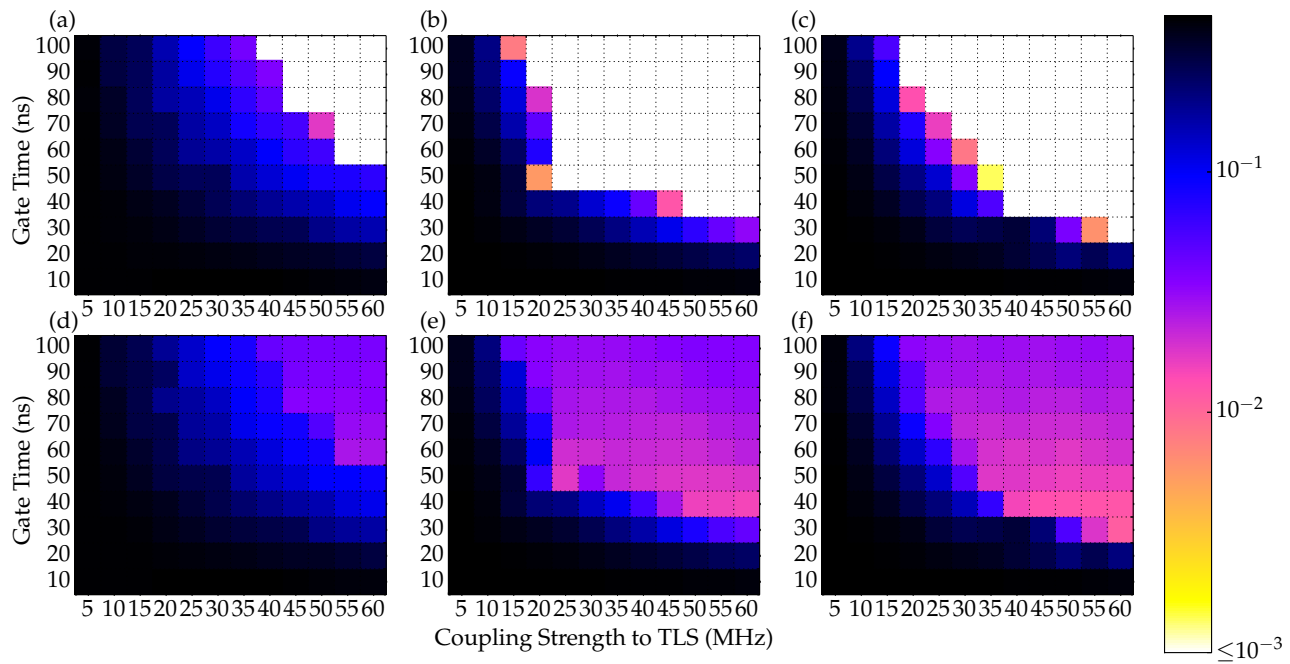}
  \caption{(color online)
    Error after optimization for $\mathrm{diag}(1,-1,1,1)$ for 
    three different anharmonicities, 
    $\beta=0\,$MHz (a,d), $40\,$MHz (b,e), $150\,$MHz (c,f)
    and infinite $T_1$ of both qudit and TLS (a-c) vs realistic $T_1$ 
    (d-f, $T_1=5\,\mu$s (qudit), $T^{(1)}_1=1\,\mu$s (TLS)). 
  }
  \label{fig:control_nodiss}
\end{figure*}

Use of the environment as a resource 
is further illustrated in Fig.~\ref{fig:control_nodiss} which explores
the dependence of the best possible error on qudit anharmonicity
and coupling strength between qudit and TLS: For very small coupling
no solution can be found and the error remains of the order one. 
On the other hand,
a single, only moderately coupled TLS in the primary bath 
is sufficient to yield good fidelities even for weak or zero
anharmonicity. In the latter case (Fig.~\ref{fig:control_nodiss}(a,d)),   
the desired diagonal unitaries can be
realized if the operation time is sufficiently long. This 
can only be exploited
for good $T_1$ times, utilizing the level-dependent coupling strengths.
The control problem becomes much easier for non-zero anharmonicity, 
with a subtle interplay between the requirements of resolving the
qudit levels and sufficient interaction with all qudit levels. 
The latter corresponds to small anharmonicity
(Fig.~\ref{fig:control_nodiss}(b,e)) and subsequently allows good
results even for weak coupling, whereas energy resolution is best for
larger anharmonicity (Fig.~\ref{fig:control_nodiss}(c,f)), which in
turn allows for very short operation times.
For fixed anharmonicity, one expects larger
coupling strengths and longer gate times to allow for better
fidelities. A few exceptions to this rule, which are observed 
in Fig.~\ref{fig:control_nodiss}, can be attributed to the
numerical nature of our controllability analysis.
The observation that for a very weakly coupled TLS there exists 
no anharmonicity and no gate time that lead to even moderate
fidelities is clear evidence that the primary bath TLS is 
essential for the generation of arbitrary diagonal unitaries.

\begin{table}
  \centering
  \begin{tabular}{|c|c|c|c|}
    \hline
    $\Delta^{(2)}$  &$S^{(2)}$                 &$T^{(2)}_1$     &
    error                    \\ 
    \hline
    50 MHz        &40 MHz                  &2000 ns       & $3.076
    \cdot 10^{-2}$          \\ 
    50 MHz        &40 MHz                  &200 ns        & $4.052
    \cdot 10^{-2}$          \\ 
    50 MHz        &40 MHz                  &40 ns         & $7.867
    \cdot 10^{-2}$          \\ \hline
    50 MHz        &10 MHz                  &2000 ns       & $3.196
    \cdot 10^{-2}$          \\ 
    50 MHz        &10 MHz                  &200 ns        & $3.564
    \cdot 10^{-2}$          \\ 
    50 MHz        &10 MHz                  &40 ns         & $4.241
    \cdot 10^{-2}$          \\ \hline
    450 MHz       &40 MHz                  &2000 ns       & $1.659
    \cdot 10^{-2}$          \\ 
    450 MHz       &40 MHz                  &200 ns        & $1.652
    \cdot 10^{-2}$          \\ 
    450 MHz       &40 MHz                  &40 ns         & $1.758
    \cdot 10^{-2}$          \\ \hline
    450 MHz       &10 MHz                  &2000 ns       & $1.663
    \cdot 10^{-2}$          \\ 
    450 MHz       &10 MHz                  &200 ns        & $1.674
    \cdot 10^{-2}$          \\ 
    450 MHz       &10 MHz                  &40 ns         & $1.675
    \cdot 10^{-2}$          \\ 
    \hline
  \end{tabular}
  \caption{Error after optimization for $\mathrm{diag}(1,-1,1,1)$
    with two primary bath TLS (parameters for qudit and first TLS as in 
    Fig.~\ref{fig:controls}, second TLS positioned $\Delta^{(2)}$ below
    $\omega^{(1)}$). For comparison, the error obtained for
    a single TLS is $1.652\cdot 10^{-2}$. 
  } 
  \label{tab:MutliTLS}
\end{table}
While the primary bath may provide interactions with the 
system that can be used as a resource for control, it can also have
detrimental effects on the system, in particular when more than one
TLS comes into 
play. This is likely to happen since number, position and coupling
strength of the TLS cannot be controlled in the preparation
of the actual devices. We therefore analyze 
the presence of an additional primary bath TLS in our optimizations,
cf. Table~\ref{tab:MutliTLS}. If the TLS are not too close to each
other, a suitable control can suppress the effect of 
the additional TLS even if it is strongly coupled and very noisy. On
the other hand, and not surprisingly so, the stronger a closely lying  
second TLS is coupled to the qudit, the more difficult it is to 
maintain good fidelities.
This is due to the fact that the gate time needs to be sufficiently
long to resolve the energy difference between the two TLS. 
Adding more TLS to the primary bath does not change the picture shown
in Table~\ref{tab:MutliTLS}: In optimizations with as many as four
strongly coupled primary bath TLS, the error is increased by less than
a factor of 2 compared to the error for a single TLS 
if none of the additional TLS is close to the favourable one 
and less than a factor of 4 if a moderately lossy TLS is in its
vicinity.  

In summary, we have shown that a non-Markovian environment can be
exploited for quantum control, enabling realization of all quantum
operations in SU(4) where the system alone allows only for SO(4). The
enhanced controllability results from an
effective control over the system-bath coupling by moving the system into
and out of resonance with a selected bath mode. 
Fast implementations of this control scheme were obtained with OCT 
such that the
errors are solely $T_1$-limited. Our model and results are directly
applicable to superconducting phase and transmon circuits for which we
predict, with reasonably simple controls, errors below one per cent
for state of the art decoherence times. 

More generally, our results provide a new perspective on open quantum
systems -- the environment can act as a resource for (almost) unitary
quantum control which can be exploited using OCT to get the details of
the dynamics right. It requires one or a few environmental modes to be
sufficiently isolated and sufficiently strongly coupled to the
system. These conditions are met for a variety of solid-state
devices other than superconducting circuits, for example NV centers in
nanodiamonds or nanomechanical oscillators. In addition, 
on an abstract level, our work calls for a comprehensive investigation
of controllability of open quantum systems, in order to gain a
rigorous understanding of when and how non-Markovianity is beneficial
for quantum control. 

\begin{acknowledgments}
  We would like to thank Ronnie Kosloff for fruitful
  discussions. Financial support by the DAAD
  and the ISF (Bikura Grant No. 1567/12)
  is gratefully acknowledged. 
\end{acknowledgments}

\end{document}